
\magnification=1200
\hsize=13cm

%
%
\centerline{\bf 1. Introduction }
\vskip 1cm
Quasitriangular Hopf algebras [1-4] are currently being explored with
a view to new applications in several areas of physics [5]. Interesting
examples of this structure are deformations of classical Lie algebras
and Lie groups [1-4], where a parameter $q$, real or complex, is
introduced in such a way that in the limit $q \rightarrow 1$ one
recovers the
non-deformed structure.

There has been an intense activity in this area in the last few years
and recently an interesting connection between the quantum universal
enveloping algebra ${\cal U}_q(SU(2))$
and anyons [6-9] has been found [10]. It was shown to be
possible to realize  ${\cal U}_q(SU(2))$ by a generalized
Schwinger construction [11], using non-local, intrinsically two
dimensional objects, with braiding properties, interpolating between
{}fermionic and bosonic oscillators, defined on a lattice $\Omega$.
These anyonic oscillators are quite different from the $q$-oscillators
introduced a few years ago in order to realize the quantum enveloping
algebras ${\cal U}_q(A_r)$,
${\cal U}_q(B_r)$, ${\cal U}_q(C_r)$, ${\cal U}_q(D_r)$ [12-16]
and the quantum exceptional algebras [17], because $q$-oscillators
are local operators which can live in any dimension.

The realization of ${\cal U}_q(A_r)$ was immediately
{}found [18] using a set of $r+1$ anyonic oscillators. In this paper
we generalize this construction to all deformed classical Lie algebras.
As in references [10,18], the deformation parameter $q$ is connected to
the statistical parameter $\nu$ by $q=\exp({\rm i}\pi\nu)$
{}for ${\cal U}_q(A_r)$,
${\cal U}_q(B_r)$, ${\cal U}_q(D_r)$ and
by $q=\exp(2{\rm i}\pi\nu)$ for ${\cal U}_q(C_r)$.

A unified treatment is provided by a sort of bosonization formula which
expresses the generators of the deformed algebras in terms of the
undeformed
ones. This relation resembles the bosonization formula [19] of two
dimensional quantum field theories (QFT), which relates bosons and
{}fermions through an exponential of bosonic fields, and in the same
way looks like the anyonization of planar QFT [20].

The building blocks  of our ``bosonization formula'' are
representations of the deformed algebras on each site of the lattice,
which do not depend of the deformation parameter; this happens when
all the $SU(2)$ subalgebras relevant to the simple roots are in spin
$0$ or $1/2$ representation. The fundamental representations of all
classical algebras share this property, which for ${\cal U}_q(A_r)$,
${\cal U}_q(B_r)$, ${\cal U}_q(D_r)$
follows directly from the Schwinger construction in terms of anyons,
since these are hard core objects; for ${\cal U}_q(C_r)$ the hard core
condition must be strengthened to prevent
any two anyons, even of different kinds, from sitting on the same
site. Moreover for ${\cal U}_q(C_r)$ the anyons have to be grouped into
pairs: the two
anyons of each pair have opposite statistical parameter and
produce a phase also when they are braided with each other.

We would like to stress that our ``bosonization formula'' is
different from the relation between the generators of
quantum and classical algebras found few years
ago [5,21]. Our expression is two dimensional and non local
since it  involves an exponential of the generators of the
Cartan subalgebra weighted with the angle function defined
on the two dimensional lattice.
As discussed in reference [10], the angle
function and its relevant cuts both provide an ordering on the
lattice and allow to distinguish between clockwise and
counterclockwise braidings; therefore the whole construction
cannot be extended to higher dimensions. However we remark that
anyons can consistently be defined also on one dimensional chains;
in such a case they become local objects and their braiding
properties are dictated by their natural ordering on the line.
Consequently, the whole treatment of the present paper and
refs. [10,18] works equally well on one dimensional chains.
As pointed out in section 6, in the one dimensional case it is
possible to extend the construction also to real values of
the deformation parameter $q$.
This paper is organized as follows. In section 2, we review
briefly the main results concerning anyonic oscillators and
lattice angle function. In section 3, we
discuss the ``bosonization formula'' for the quantum version of
the classical Lie algebras. In section 4 we present the fermionic
realization of the Lie algebras of type $A_r$, $B_r$, $D_r$ and
the anyonic realization of the corresponding deformed algebras and
in section 5 we extend the procedure to the algebras of type $C_r$.
Section 6 is devoted to some final remarks.

\vskip 3cm

%
%
\centerline{\bf 2. Lattice Angle Function and
Anyonic Oscillators }
\vskip 1cm
In this section, following ref. [10], we review
the construction of anyonic
oscillators defined on a two-dimensional square
lattice $\Omega$.

Anyonic oscillators are two-dimensional non-local
operators [22-26] which
interpolate between bosonic and fermionic oscillators.
On a lattice
they can be constructed by means of the generalized
Jordan-Wigner
transformation [19] which in our case transmutes
fermionic oscillators
into anyonic ones. Its essential ingredient is the
lattice angle function  $\Theta ({\bf x},{\bf y})$
that was defined in a very
general way in ref.
[23,20]. Here we describe concisely the particular definition
of  $\Theta ({\bf x},{\bf y})$ given in ref. [10].

We begin by embedding the lattice $\Omega$ with
spacing one into a
lattice $\Lambda$ with spacing $\epsilon $, which eventually
will be sent to zero. Then to
each point ${\bf x} \in \Omega$ we associate a cut
${\gamma}_x$, made with bonds of the dual lattice $\tilde{\Lambda}$
{}from minus infinity to ${\bf x}^*={\bf x}+{\bf o}^*$ along $x$-axis,
with ${\bf o}^*=\left({\epsilon\over 2},
{\epsilon\over 2} \right)$
the origin
of the dual lattice $\tilde{\Lambda}$. We denote by
${\bf x}_{\gamma}$ the point
${\bf x} \in \Omega $ with its associated cut
${\gamma}_x$.

Given any two distinct points ${\bf x}=(x_1,x_2)$
and
${\bf y}=(y_1,y_2)$ on $\Omega $,
and their associated cuts ${\gamma}_x$ and ${\gamma}_y$,
in the limit $\epsilon \to 0$ it is possible to
show that [10]
$$
\Theta_{{\gamma}_x}({\bf x},{\bf y}) -
\Theta_{{{\gamma}_y}}({\bf y},
{\bf x}) =
\left\{\matrix{
 \pi \,{\rm sgn}(x_2-y_2)  ~~~~~
 {\rm for}~~~x_2\not=y_2 \ \ , \cr
 \pi \,{\rm sgn}(x_1-y_1)  ~~~~~
 {\rm for}~~~x_2=y_2 \ \ ,
\cr}\right.
\eqno(2.1)
$$
with $\Theta_{{\gamma}_x}({\bf x},{\bf y})$ being
the angle of the point $\bf x$ measured from the point
${\bf y}^* \in {\tilde\Lambda}$
with respect to a line parallel to the positive
$x$-axis.

Eq. (2.1) can be used to endow  the lattice with an ordering
which will be very useful
in handling anyonic oscillators. We define
${\bf x} > {\bf y}$
by choosing the positive sign
in eq. (2.1), {\it i.e.}
$$
{\bf x}>{\bf y}
\Longleftrightarrow
\left\{\matrix{
 x_2>y_2 \ \ , \cr
 x_2=y_2~,~x_1>y_1 \ \ . \cr}\right.
\eqno(2.2)
$$
{}From eqs. (2.1-2) it follows that
$$
\Theta_{{\gamma}_x}({\bf x},{\bf y}) -
\Theta_{{{\gamma}_y}}({\bf y},
{\bf x}) = \pi
{}~~~~~{\rm for}~~~{\bf x}>{\bf y}\ \ .
\eqno(2.1')
$$
Even if unambiguous, this definition of the angle
$\Theta({\bf x},{\bf y})$ is not
unique since it depends on the choice of the cuts.
Suppose now, instead of choosing ${\gamma}_x$,
we choose for each point of the lattice a
cut ${\delta}_x$ made with
bonds of the dual lattice $\tilde\Lambda $
{}from plus infinity to $^*{\bf x}$
along $x$-axis,
 with $^*{\bf x}={\bf x}-{\bf o}^*$ .
In this case it can be shown that the relation between
 the angle of two distinct points ${\bf x},
 {\bf y} \in \Omega$
 becomes [10]
$$
\tilde\Theta_{{\delta}_x}({\bf x},{\bf y}) -
\tilde\Theta_{{{\delta}_y}}({\bf y},{\bf x})
=\left\{\matrix{
 -\pi \,{\rm sgn}(x_2-y_2)  ~~~~~{\rm for}~~~x_2\not
=y_2 \ \ , \cr
 -\pi \,{\rm sgn}(x_1-y_1)  ~~~~~{\rm for}~~~x_2
=y_2 \ \ . \cr}\right.
\eqno(2.3)
$$
Notice that $\tilde\Theta_{{\delta}_x}({\bf x},{\bf y})$
 is now the angle of {\bf x} as seen from $^*{\bf y} \in
{\tilde\Lambda}$
with respect to a line parallel to the negative
$x$-axis.

The choice of the cuts $\delta_x$ would therefore
induce an opposite order with respect
to the one defined in (2.2). Keeping instead the
ordering (2.2), eq. (2.3) reads
$$
\tilde\Theta_{{\delta}_x}({\bf x},{\bf y}) -
\tilde\Theta_{{{\delta}_y}}({\bf y},{\bf x})
=- \pi
  ~~~~~{\rm for}~~~{\bf x}>{\bf y}\ \
\eqno(2.3')
$$
We can also have the relation between $\Theta_{\gamma}$
and $\tilde\Theta_{\delta}$. Using their definitions
we get [10]
$$
\tilde\Theta_{{\delta}_x}({\bf x},{\bf y}) -
\Theta_{{{\gamma}_x}}({\bf x},{\bf y})
=\left\{\matrix{
 -\pi \, ~~~~~{\rm for}~~~{\bf x}>{\bf y}\ \ , \cr
{}~~ \pi \, ~~~~~~{\rm for}~~~{\bf x}<{\bf y}
\ \ ,  \cr}\right.
\eqno(2.4)
$$
and using (2.1') and (2.4) it follows that
$$
\tilde\Theta_{{\delta}_x}({\bf x},{\bf y}) -
\Theta_{{{\gamma}_y}}({\bf y},{\bf x})
=0 ~~~~~~~\forall~~
{\bf x},{\bf y}\ \ .
\eqno(2.5)
$$

We are going to use now the angle functions
$\Theta_{{{\gamma}_x}}({\bf x},{\bf y})$ and
$\tilde\Theta_{{\delta}_x}({\bf x},{\bf y}) $
to define two kinds of parity related anyonic oscillators.
We define anyonic oscillators of type $\gamma$ and $\delta$
as follows
$$
a_{i}({\bf x}_{\alpha})=K_{i}({\bf x}_{\alpha})~c_{i}({\bf x})
 ~~~~~~({\rm no~ sum~ over~} i)
\eqno(2.6)
$$
with $\alpha_{x} = \gamma_{x}$ or $ \delta_{x}$,  $i=1,\cdots,N$;
the disorder operators $K_{i}({\bf x}_{\alpha})$
[19, 27] are given by
$$
\eqalign{
K_i({\bf x}_{\alpha}) &= {\rm \exp}{\Big[{\rm i} \,
\nu\sum\limits_{{{\bf y}\in \Omega}\atop {{\bf y}\not = {\bf x}}}
\Theta_{{{\alpha}_x}}({\bf x},{\bf y})~
\big(n_i({\bf y})- {1\over 2}\big)}\Big]}
\eqno(2.7)
$$
$$
n_i({\bf y})=c^\dagger_i({\bf y})c_i({\bf y}) \  ;
\eqno(2.8)
$$
$\nu$ is the statistical parameter and $c_i({\bf x})$,
$c^\dagger_i({\bf x})$
are fermionic oscillators defined on $\Omega$ obeying the
usual anticommutation relations
$$
\eqalign{
\Big\{c_i({\bf x})~,~c_j({\bf y})\Big\}&=0\ \ ,
\cr
\Big\{c_i({\bf x})~,~c^\dagger_j({\bf y})\Big\}&=
\delta_{ij}~\delta({\bf x},{\bf y})\,
\ \ }
\eqno(2.9)
$$
where
$$
\delta({\bf x},{\bf y})=\left\{\matrix{
1  ~~~~~{\rm if}~~~
{\bf x}={\bf y} \ \ {}\cr
{}~0  ~~~~~{\rm if}~~~
{\bf x}\not={\bf y}\ \ . \cr}\right.
\eqno(2.10)
$$

We remark that the disorder operator $K_i({\bf x}_{\alpha})$
differs from the one defined in [10]  because of the
subtraction of the background term $1\over 2$ from the
{}fermion occupation number $n_i({\bf y})$. This does not
change the result of refs. [10, 18] for
${\cal U}_q(A_r)$, but is crucial for
${\cal U}_q(B_r)$ and ${\cal U}_q(D_r)$.
\par
Using (2.1') and (2.9) we get the following generalized commutation
relations  for   anyonic oscillators of type $\gamma$
$$
\eqalignno{
a_i({\bf x}_{\gamma})\,a_i({\bf y}_{\gamma})+&\,
q^{-1}a_i({\bf y}_{\gamma})\,a_i({\bf x}_{\gamma})=0 \ \ ,
&(2.11{\rm a})\cr
a_i({\bf x}_{\gamma})\,a^\dagger_i({\bf y}_{\gamma})+&\,
q~~~a^\dagger_i({\bf y}_{\gamma})\,a_i({\bf x}_{\gamma})=0 \ \ ,
&(2.11{\rm b})
}
$$
{}for ${\bf x}>{\bf y}$
and $q=\exp({\rm i}\pi\nu)$.
If ${\bf x}={\bf y}$ we have
$$
\left(a_i({\bf x}_\gamma)\right)^2=0 \ \ ,
\eqno(2.12{\rm a})
$$
$$
a_i({\bf x}_{\gamma})~a^\dagger_i({\bf x}_{\gamma})+
a^\dagger_i({\bf x}_{\gamma})~a_i({\bf x}_{\gamma})=1 \ \ .
\eqno(2.12{\rm b})
$$
Eqs. (2.11-12) mean that anyonic oscillators are hard core
objects
and obey $q$-commutation relations at different points of the
lattice
but standard anticommutation relations at the
same point.\footnote{*}{ Here and in the following we do not write
the other generalized commutation relations which can be obtained
by hermitean
conjugation, taking into account that $q^{*}=q^{-1}$. }

Of course different oscillators obey the ordinary
anticommutation relations
$$
\eqalign{
\Big\{a_i({\bf x}_\gamma)~,~a_j({\bf y}_\gamma)\Big\}=
\Big\{a_i({\bf x}_\gamma)~,~a^\dagger_j
({\bf y}_\gamma)\Big\}=0\ \ ,\cr
\forall~~{\bf x},\,{\bf y}\in\Omega ~~~~{\rm and}
{}~~\forall~~i,j~=1,\cdots,N ,~~i\not=j \ \ .
}
\eqno(2.13)
$$

The commutation relations among anyonic oscillators of
type $\delta$ can be obtained from the previous ones,
(2.11-13), by replacing $q$ by $q^{-1}$ and $\gamma$
by $\delta$. This is due to the fact that
$\delta$ ordering can be obtained from
$\gamma$ ordering by a parity transformation which,
 as is well known, changes the braiding phase $q$ into
$q^{-1}$ (see for istance [9]).

To complete our discussion
we compute the commutation relations between type $\gamma$
and type $\delta$
oscillators. By using  eqs. (2.4-6) one gets
$$
\eqalignno{
\Big\{a_i({\bf x}_{\delta})~,~a_j({\bf y}_{\gamma})
\Big\}&=\,
0~~~~~~~~~~~\forall~~{\bf x},\,{\bf y} \ \ , \ \ \forall~~i,\,j \ \,
&(2.14{\rm a})\cr
\Big\{a_i({\bf x}_{\delta})~,~a^\dagger_j({\bf y}_{\gamma})
\Big\}&=\,
\delta_{ij}~\delta({\bf x},{\bf y})\ \
q^{-\Big[\sum\limits_{{\bf z}<{\bf x}}-
\sum\limits_{{\bf z}>{\bf x}}\Big]
\big(n_i({\bf z}) - {1\over 2} \big)} \ \,
&(2.14{\rm b})
}
$$

It should be clear from the previous discussion that anyonic
oscillators do not have anything to do with $q$-oscillators
introduced a few years ago (ref. [12,13]). The main reason is
that the generalized commutation relations (2.11-14)
are meaningful only on an ordered lattice. Ordering is
natural on a linear chain, where eqns. (2.11-14) could be
postulated  a priori, defining  one-dimensional
``local anyons''. Instead on a two dimensional lattice,
ordering follows from the introduction
of an angle function with its associated cut.
In such a case oscillators are non-local objects, contrarily
to the deformed $q$-oscillators which are local
and can be defined in any dimension.
\vskip 3cm

%
%

\centerline{\bf 3. A Bosonization Formula for
Quantum Algebras}
\vskip 1cm

By construction, the deformed Lie algebras reduce to the
undeformed ones when the deformation parameter $q$ goes to $1$.
When $G$ is a classical Lie algebra the connection is even closer:
there exists a set of non trivial representations of ${\cal U}_q(G)$
which do not depend on $q$ and therefore are common to the deformed
and undeformed enveloping algebras.\footnote{*}{Actually this property
holds also for $E_6$ and $E_7$, but not for the other exceptional
algebras. The whole discussion of this section can  thus be
referred also to ${\cal U}_q(E_6)$ and  ${\cal U}_q(E_7)$.}
This happens when all the $SU(2)$ subalgebras relevant to the simple
roots are in spin $0$ or $1/2$ representation; we call ${\Re}_{(0,1/2)}$
the set of representations with this property.

Another important fact is that the fundamental representations of
classical Lie algebras, listed in fig.1 [28], belong to the set
${\Re}_{(0,1/2)}$; by fundamental representation we mean an
irreducible representation such that any other representation
can be constructed
{}from it by taking tensor products, or, equivalently, by repeated use
of comultiplication. For these reasons it is possible to express the
generators of the $q$-deformed Lie algebras
in terms of the generators of the undeformed algebras in a
{}fundamental representation.

The plan of this section is the following: at first we show that
the representations of ${\cal U}_q(G)$ belonging to the set
${\Re}_{(0,1/2)}$ do not depend on the deformation parameter $q$;
then we write the ``bosonization formula''
which expresses the generators of ${\cal U}_q(G)$
by means of an exponential involving
the undeformed generators on each site of a two dimensional
lattice and the angle functions, $\Theta({\bf x},{\bf y})$,
defined in Sect. 2.

The generalized commutation relations of
${\cal U}_q(G)$  in the Chevalley basis are
$$
\eqalignno{
\Big[H_I~&,~H_J\Big]=
0 \ \ ,
&(3.1{\rm a})\cr
\Big[H_I~&,~E_J^{\pm}\Big]=
\pm a_{IJ}~E_J^{\pm} \ \ ,
&(3.1{\rm b})\cr
\Big[E_I^+~&,~E_J^-\Big]=
\delta_{IJ}~\Big[H_I \Big]_{q^{}_I}\ \ ,
&(3.1{\rm c})\cr
\sum_{\ell=0}^{1-a_{IJ}}(-1)^\ell&{{1-a_{IJ}}\brack \ell}_{q^{}_I}
\left(E_{I}^{\pm}\right)^{1-a_{IJ}-\ell}~E_{J}^{\pm}
{}~\left(E_{I}^{\pm}\right)^\ell=0 \ \ ,
&(3.1{\rm d}) \cr}
$$
where $H_I$ are the generators
of the Cartan subalgebra, $E_I^{\pm}$
are the step operators corresponding
to the simple root $\alpha_I$
and $a_{IJ}$ denotes the Cartan matrix, {\it i.e.}
$$
a_{IJ}=\left\langle \alpha_I,\alpha_J \right\rangle =
2~{{(\alpha_I,\alpha_J)}\over{(\alpha_I,\alpha_I)}} \ \
{}~~~~~~I,J =1,2,...r \ , \, r=rank(G) \,
\eqno(3.2)
$$
In eqs. (3.1) we have used the notations
$$
\eqalign{
{[x]}_q&={{q^x-q^{-x}}\over{q-q^{-1}}}\ \ ,\cr
{{m\brack n}}_q&={{{[m]}_q!}\over {{[m-n]}_q!~{[n]}_q!}} \ \ ,\cr
{}\cr
{[m]}_q!&={[m]}_q{[m-1]}_q\cdots{[1]}_q\ \ .}
\eqno(3.3)
$$
where $q$ is the deformation parameter. Moreover, $q^{}_I$ is defined
as
$$
q^{}_I~=~q^{{1\over{2}}{(\alpha_I,\alpha_I)}},
\eqno(3.4)
$$
so that
$$
{q^{}_I}^{a_{IJ}}~=~{q_J^{}}^{a_{JI}}.
\eqno(3.5)
$$
To complete the definition of ${\cal U}_q(G)$,
the comultiplication $\Delta$,
the antipode $S$ and the co-unit $\epsilon$ are given by
$$
\eqalign{
\Delta(H_I) &=\,
 H_I \otimes{\bf 1}+{\bf 1}\otimes H_I \ \ ,
\cr
\Delta(E_I^{\pm})&=\,
E_I^{\pm}\otimes {q{}_I}^{{H_I}/2}+
{q{}_I}^{-{H_I}/2}\otimes E_I^{\pm}\ \ ,
\cr
S({\bf 1}) &=\,
{\bf 1}~,~S(H_I)=-H_I \ \ ,
\cr
S(E_I^{\pm })&=\,
-{q{}_I}^{{H_I}/2} E_I^{\pm}{q{}_I}^{-{H_I}/2} \ \ ,
\cr
\epsilon({\bf 1}) &=\,
 1 ~,~ \epsilon(H_I) =\epsilon(E_I^{\pm})=0  \ \ .
}
\eqno(3.6)
$$
Let us now denote by $h_I$ and $e_I^{\pm}$ the generators
$H_I$ and $E_I^{\pm}$ in a representation belonging to the
set ${\Re}_{(0,1/2)}$; then:
\item{$i)$} the eigenvalues of $h_I$,  {\it i.e.} the Dynkin
labels of any weight, can be only
$0$ or $\pm 1$, and, equivalently
\item{$ii)$}$(e_I^{\pm})^2~=~0$.

Therefore, for any value
of $q$, due to the definition (3.3) and
the property $i)$,
$$
\big[h_I \big]_{q^{}_I}~=~ h_I\ \ .
\eqno(3.7)
$$
Moreover the deformed Serre relation (3.1d),
which reads
$$
{}~~~~~~~~~~~~\Big[E_I^{\pm}~,~E_J^{\pm}\Big]=0
{}~~~~~~~~~~{\forall}~~ I,J~/~ a_{IJ}=~0
\eqno(3.8)
$$
becomes, due to the property $ii)$
$$
-(q^{}_I~+~q_I^{-1})~e_I^{\pm}~e_J^{\pm}~e_I^{\pm}=0
{}~~~~~~~~~~{\forall}~~ I,J~/~a_{IJ}=-1\ \ ,
\eqno(3.9)
$$
and is identically satisfied for $I,J$ such that $a_{IJ}=-2$.
\par
This shows that, for the representations in ${\Re}_{(0,1/2)}$,
the deformed commutation relations (3.1)
are independent of the deformation parameter $q$ and
therefore coincide with the undeformed ones. Thus
the deformed and the undeformed classical Lie algebras share the
same fundamental representations, because they belong
to the set ${\Re}_{(0,1/2)}$.
\par
All the other representations can be obtained from a
{}fundamental one by repeated use of comultiplication; the
difference between ordinary and deformed Lie algebras
is just in the different rules of comultiplication.
\par

To make contact with sect.2, we assign a fundamental
representation to each point ${\bf x}$
of an ordered two-dimensional (or one-dimensional)
lattice $\Omega$;
the local generators satisfy the following generalized
commutations relations:
$$
\eqalignno{
\Big[h_I({\bf x})~&,~h_J({\bf y})\Big]=
0 \ \ ,
&(3.10{\rm a})\cr
\Big[h_I({\bf x})~&,~e_J^{\pm}({\bf y})\Big]=
\pm \delta({\bf x},{\bf y})~ a_{IJ}~e_J^{\pm}({\bf x}) \ \ ,
&(3.10{\rm b})\cr
\Big[e_I^{+}({\bf x})~&,~e_J^{-}({\bf y})\Big]=
\delta({\bf x},{\bf y})~\delta_{IJ}~
\big[h_I({\bf x})\big]_{q^{}_I} \ \ ,
&(3.10{\rm c})\cr
\sum_{\ell=0}^{1-a_{IJ}}(-1)^\ell&{{1-a_{IJ}}
\brack \ell}_{q^{}_I}
\left(e_I^{\pm}({\bf x})\right)^{1-a_{IJ}-\ell}
{}~e_I^{\pm}({\bf x})
{}~\left(e_I^{\pm}({\bf x})\right)^\ell=0 \ \ ,
&(3.10{\rm d})\cr
\Big[e_I^{\pm}({\bf x})~&,~e_J^{\pm}({\bf y})\Big]=
0 ~~~~{\rm for}~~~{\bf x}\not={\bf y}\ \ .
&(3.10{\rm e})
}
$$
{}From the previous discussion it should be clear that
the relations (3.10)
are just formally ``deformed'', as
the fundamental representations of classical Lie
algebras belong to the set ${\Re}_{(0,1/2)}$;
nevertheless, writing them as deformed commutation relations
will be useful for our discussion.

In fact the iterated coproduct for the deformed enveloping algebra reads
$$
H_I =\sum_{{\bf x}\in \Omega} H_I({\bf x})
{}~~~~,~~~~
E_I^{\pm} =
\sum_{{\bf x}\in \Omega}E_I^{\pm}({\bf x})
\eqno(3.11)
$$
where
$$
\eqalignno{
H_I({\bf x}) &=\prod_{{\bf y}<{\bf x}}{}^{\!\!\otimes}~
{\bf 1}_{\bf y}\otimes h_I({\bf x})\otimes
\prod_{{\bf z}>{\bf x}}{}^{\!\!\otimes}~{\bf 1}_{\bf z}\ \ ,
&(3.12{\rm a})
\cr
E_I^{\pm}({\bf x}) &=\prod_{{\bf y}<{\bf x}}{}^{\!\!\otimes}~
{q^{}_I}^{-{1\over2}h_I({\bf y})} \otimes
e_I^{\pm}({\bf x})\otimes
\prod_{{\bf z}>{\bf x}}{}^{\!\!\otimes}~{q^{}_I}^{{1\over2}h_I({\bf z})}\ \ ,
&(3.12{\rm b})
}
$$
and we know that consistency between product and coproduct implies
that the generators $H_I$ and $E_I^{\pm}$ defined in eqs. (3.11-12)
satisfy eqs. (3.1), once that $h_I({\bf x})$ and $e_I^{\pm}({\bf x})$
satisfy  eq. (3.10).
By checking this explicitly, we can obtain an expression equivalent to
eq. (3.12b) but more useful in this context, as follows.

The check is trivial for eqs. (3.1a) and (3.1b);
to check eq. (3.1c) one needs at first the relation
$$
\Big[E_I^{+}({\bf x})~,~E_J^{-}({\bf y})\Big]=\,
\delta({\bf x},{\bf y})~\delta_{IJ}~
\prod_{{\bf w}<{\bf x}}{}^{\!\!\otimes}~{q^{}_I}^{-h_I({\bf w})}
\otimes
\big[h_I({\bf x})\big]_{q^{}_I}  \otimes
\prod_{{\bf z}>{\bf x}}{}^{\!\!\otimes}~{q^{}_I}^{h_I({\bf z})}\ \ ,
\eqno(3.13)
$$
which follows from the definition (3.12b), from the
commutation relations (3.10b-c) and from the
identity (3.5); then one can complete the proof
by complete induction, following ref. [10].
\par
{}Finally the deformed Serre relation follows from eq.(3.10d)
and from the braiding relations
$$
E_I^{\pm}({\bf x})E_J^{\pm}({\bf y})=\left\{\matrix{
{q^{}_I}^{\pm a_{IJ}}~~ E_J^{\pm}({\bf y})\,E_I^{\pm}({\bf x})
{}~~~~~{\rm for}~~~
{\bf x}>{\bf y}\ \ , \cr
{}\cr
{q^{}_I}^{\mp a_{IJ}}~~ E_J^{\pm}({\bf y})\,E_I^{\pm}({\bf x})
{}~~~~~{\rm for}~~~
{\bf x}<{\bf y}\ \  \cr}\right.
\eqno(3.14)
$$
which are a consequence of the definitions of (3.12b) and the
commutation relations (3.10b,e).
Let us now introduce a new set of non local densities
$H_I({\bf x})$,
$E_I^{\pm}({\bf x})$ defined using the angles
$\Theta_{{\gamma}_x}({\bf x},{\bf y})$ and
$\tilde\Theta_{{\delta}_x}({\bf x},{\bf y})$
discussed in section 2:
$$
\eqalignno{
H_I({\bf x}) &=\prod_{{\bf y}<{\bf x}}{}^{\!\!\otimes}
{}~{\bf 1}_{\bf y}\otimes h_I({\bf x})
\prod_{{\bf z}>{\bf x}}{}^{\!\!\otimes}~{\bf 1}_{\bf z}\ \ ,
&(3.15{\rm a})
\cr
E_I^{+}({\bf x}) &=e_I^{+}({\bf x})
\otimes \prod_{{\bf y}\not={\bf x}}{}^{\!\!\otimes}
{}~{q^{}_I}^{-{1\over\pi} \Theta_{{\gamma}_x}({\bf x},
{\bf y})\,h_I({\bf y})}\ \ ,
&(3.15{\rm b})
\cr
E_I^{-}({\bf x}) &=e_I^{-}({\bf x}) \otimes
\prod_{{\bf y}\not={\bf x}}{}^{\!\!\otimes}
{}~{q^{}_I}^{
{1\over\pi} \tilde\Theta_{{\delta}_x}({\bf x},
{\bf y})\,h_I({\bf y})}\ \ .
&(3.15{\rm c})
}
$$
Using the properties of
$\Theta_{{\gamma}_x}({\bf x},{\bf y})$ and
$\tilde\Theta_{{\delta}_x}({\bf x},{\bf y})$ given
by eqs. (2.1'), (2.3') and (2.5), it is possible to
show that $E_I^{\pm}({\bf x})$ have exactly the same commutation
and braiding relations as the operators defined in (3.12b), that is
eqs. (3.13) and (3.14) hold exactly as in the previous case.

It is thus obvious that the global generators
$H_I$ and $E_I^\pm$ obtained by inserting the
densities (3.15) instead of (3.12) into eqs. (3.11) still satisfy
the deformed algebra of ${\cal U}_q(G)$.

It is interesting to observe that the new generators (3.15)
can be introduced also for a one dimensional lattice.
In that case it is enough to define,
consistently with eqs. (2.1'), (2.3') and (2.5),
$$
\Theta_{{{\gamma}_x}}({\bf x},{\bf y})
=\left\{\matrix{
+{\pi\over 2}  ~~~~~{\rm for}~~~
{\bf x} > {\bf y} \ \ \cr
-{\pi\over 2}  ~~~~~{\rm for}~~~
{\bf x} < {\bf y}
\ \ ,  \cr}\right.
\eqno(3.16{\rm a})
$$
and
$$
\tilde\Theta_{{\delta}_x}({\bf x},{\bf y})
=\left\{\matrix{
-{\pi\over 2}  ~~~~~{\rm for}
{}~~~{\bf x} > {\bf y}\ \ , \cr
+{\pi\over 2}  ~~~~~{\rm for}
{}~~~{\bf x} < {\bf y}
\ \ .  \cr}\right.
\eqno(3.16{\rm b})
$$
to make eqs.(3.15) coincide with the iterated coproduct (3.12).

As the fundamental representations of classical Lie algebras
belong to the set ${\Re}_{(0,1/2)}$ the generators
$h_I$ and $e_I^{\pm}$ can be considered as generators both
of the deformed and the undeformed algebras.
Therefore, on a one or two dimensional lattice,
the ``bosonization formula'' (3.11) and (3.15) actually gives
the generators of the deformed Lie algebras  in any representation
in terms of the undeformed ones in the fundamental representation.

\vskip 3cm
%
%

\centerline{\bf 4. Anyonic Construction
of ${\cal U}_q(A_r)$,
${\cal U}_q(B_r)$ and ${\cal U}_q(D_r)$}
\vskip 1cm

In this section we are going to show that
eqs. (3.15) for the algebras ${\cal U}_q(A_r)$,
${\cal U}_q(B_r)$ and ${\cal U}_q(D_r)$ can be naturally
written in
terms of anyons.

It is well known that the classical Lie
algebras $A_r$, $B_r$ and $D_r$ can be constructed
\`a la Schwinger in terms of fermionic oscillators;
we perform this
construction on each site of the lattice $\Omega$
by using the oscillators $c_i({\bf x})$ ($i=1,2,...,N$)
with the usual anticommutation relations (2.9).
{}For the algebra
$A_r$ one needs $N=r+1$ oscillators so that
$$
\eqalign{
e_I^+({\bf x})&=c_I^\dagger({\bf x})
\,c_{I+1}({\bf x})\ \ ,\cr
e_I^-({\bf x})&=c_{I+1}^\dagger({\bf x})
\,c_I({\bf x})\ \ ,\cr
h_I({\bf x})&=n_I({\bf x})
-n_{I+1}({\bf x})\ \ ,}
\eqno(4.1)
$$
where $I=1,2,...,r$. For the algebras $B_r$
and $D_r$ instead, one needs $N=r$ oscillators.
In particular for $B_r$ we have
$$
\eqalign{
e_j^+({\bf x})&=c_j^\dagger({\bf x})
\,c_{j+1}({\bf x})\ \ ,\cr
e_j^-({\bf x})&=c_{j+1}^\dagger({\bf x})
\,c_j({\bf x})\ \ ,\cr
h_j({\bf x})&=n_j({\bf x})-
n_{j+1}({\bf x})\ \ ,}
\eqno(4.2{\rm a})
$$
{}for $j=1,2,...,r-1$ and
$$
\eqalign{
e_r^+({\bf x}) &= c_r^\dagger({\bf x})~{\cal S}({\bf x})
\ \ ,\cr
e_r^-({\bf x}) &= c_r({\bf x})~{\cal S}({\bf x})  \ \ ,\cr
h_r({\bf x})&=2n_r({\bf x})-1 \ \ ,}
\eqno(4.2{\rm b})
$$
where
$$
{\cal S}({\bf x})=\prod_{{\bf y}<{\bf x}}\prod_{I=1}^r
(-1)^{n_I({\bf y})}
\ \
\eqno(4.2{\rm c})
$$
is a sign factor introduced to make the generators
commute at different points of the lattice (cf eq. (3.10e)).
{}For the algebra $D_r$ again we have
$$
\eqalign{
e_j^+({\bf x})&=c_j^\dagger({\bf x})
\,c_{j+1}({\bf x})\ \ ,\cr
e_j^-({\bf x})&=c_{j+1}^\dagger({\bf x})
\,c_j({\bf x})\ \ ,\cr
h_j({\bf x})&=n_j({\bf x})-
n_{j+1}({\bf x})\ \ ,}
\eqno(4.3{\rm a})
$$
{}for $j=1,2,...,r-1$ and
$$
\eqalign{
e_r^+({\bf x})&=c_r^\dagger({\bf x})
\,c_{r-1}^\dagger({\bf x})\ \ ,\cr
e_r^-({\bf x})&=c_{r-1}({\bf x})
\,c_r({\bf x})\ \ ,\cr
h_r({\bf x})&=n_{r-1}({\bf x})+
n_r({\bf x})-1\ \ .}
\eqno(4.3{\rm b})
$$
It is a very easy task to check that the
generators $h_I$ and $e_I^\pm$ defined
in this way satisfy
the commutation relations (3.10) with the
appropriate Cartan matrices (see Tab. 1).
Moreover one realizes that properties $i)$ and $ii)$ of
section 3 hold: all step operators
$e_I^\pm({\bf x})$
have a vanishing square and the eigenvalues of the
Cartan generators $h_I({\bf x})$
can only be either 0 or $\pm1$.  For sake of
completeness we list in
Tab. 2 the highest weight vectors corresponding to the
{}fundamental representations of Fig. 1 and in Tab. 3 the
relevant basis vectors in the Fock space generated by the
{}fermionic operators $c_i^\dagger({\bf x})$.

Obviously all
representations can be obtained by a repeated use
of the coproduct, that is by summing over all sites
of the lattice
$$
H_I=\sum_{{\bf x}\in \Omega}h_I({\bf x}) ~~~~,~~~~
E_I^\pm=\sum_{{\bf x}\in \Omega}e_I^\pm({\bf x})\ \ ;
\eqno(4.4)
$$
according to common use here and in
the following we will always drop the symbol
$\otimes$ of the direct product.

The deformed algebras can be
obtained in exactly the same way if the
{}fermionic oscillators in eqs. (4.1), (4.2) and
(4.3) are replaced by anyonic ones.
More precisely, following [10], we write the
raising operators $E_I^+({\bf x})$
in terms of the anyonic oscillators
$a_i({\bf x}_\gamma)$
and the lowering operators $E_I^-({\bf x})$
in terms of the anyonic oscillators
$a_i({\bf x}_\delta)$ defined in eq.
(2.6). The Cartan generators
can be written using either $a_i({\bf x}_\gamma)$ or
$a_i({\bf x}_\delta)$ because
$$
a_i^\dagger({\bf x}_\gamma)
\,a_i({\bf x}_\gamma)=a_i^\dagger({\bf x}_\delta)
\,a_i({\bf x}_\delta)=c_i^\dagger({\bf x})\,c_i({\bf x})
 = n_i({\bf x}) \ \ .
\eqno(4.5)
$$
In this way for all roots of $A_r$, for the long
roots of $B_r$ and all
roots of $D_r$ but $\alpha_r$, one gets
$$
\eqalign{
E_j^+({\bf x}) &= a_j^\dagger({\bf x}_\gamma)\,
a_{j+1}({\bf x}_\gamma)\cr
&=c_j^\dagger({\bf x})\,c_{j+1}({\bf x})~
{\rm e}^{-{\rm i}\nu\sum\limits_{{\bf y}
\not={\bf x}}
\Theta_{{{\gamma}_x}}({\bf x},{\bf y})
{}~\big(n_j({\bf y})-n_{j+1}({\bf y})\big)}
\cr
&=e_j^+({\bf x})~{\rm e}^{-{\rm i}
\nu\sum\limits_{{\bf y}\not={\bf x}}
\Theta_{{{\gamma}_x}}({\bf x},
{\bf y})~h_j({\bf y})} \ \ , \cr
{}\cr
H_j({\bf x})&=n_j({\bf x}) -
n_{j+1}({\bf x}) \ \ .}
\eqno(4.6)
$$
For the short root of $B_r$ one has
$$
\eqalign{
E_r^+({\bf x}) &=a_{r}^\dagger({\bf x}_\gamma)~
{\cal S}({\bf x})~=~
c_{r}^\dagger({\bf x})
{}~{\rm e}^{-{\rm i}{\nu\over 2}
\sum\limits_{{\bf y}\not={\bf x}}
\Theta_{{{\gamma}_x}}({\bf x},{\bf y})
\big(2n_r({\bf y})-1\big)}
{}~{\cal S}({\bf x})\cr
&=e_r^+({\bf x})~{\rm e}^{-{\rm i}
{\nu\over 2}\sum\limits_{{\bf y}\not={\bf x}}
\Theta_{{{\gamma}_x}}({\bf x},{\bf y})
\,h_r({\bf y})}\cr
{}\cr
H_r({\bf x})&=2\,n_r({\bf x}) - 1
}\eqno(4.7)
$$
Finally for $\alpha_r$ of $D_r$ one has
$$
\eqalign{
E_r^+({\bf x}) &= a_r^\dagger({\bf x}_\gamma)
\,a_{r-1}^\dagger({\bf x}_\gamma)\cr
&=c_r^\dagger({\bf x})\,c_{r-1}^\dagger({\bf x})~
{\rm e}^{-{\rm i}\nu\sum\limits_{{\bf y}
\not={\bf x}}
\Theta_{{{\gamma}_x}}({\bf x},{\bf y})
\big(n_r({\bf y})+n_{r-1}({\bf y})-1\big)}
\cr
&=e_r^+({\bf x})~{\rm e}^{-{\rm i}
\nu\sum\limits_{{\bf y}\not={\bf x}}
\Theta_{{{\gamma}_x}}({\bf x},{\bf y})\,h_r({\bf y})}
\cr
{}\cr
H_r({\bf x}) &= n_r({\bf x})+n_{r-1}({\bf x})-1 \ \ .
}
\eqno(4.8)
$$
One easily realizes that eqs. (4.6), (4.7)
and (4.8) coincide with the ``bosonization formulas''
(3.15a-b) if the identification
$$
q={\rm e}^{{\rm i}\nu \pi}
\eqno(4.9)
$$
is made. Therefore $q_j=q$ for the long roots
and $q_r=q^{1/2}$ for the short root of
$B_r$.
Similarly the lowering operators are
$$
\eqalignno{
E_j^-({\bf x}) &= a_{j +1}^\dagger({\bf x}_\delta)
\,a_j ({\bf x}_\delta)~~~~j =1,..r~{\rm for}~ A_r ,
{}~~j =1...r-1~{\rm for}~B_r~{\rm and}~D_r ,
&(4.6')\cr
E_r^-({\bf x}) &=a_{r}({\bf x}_\delta)~
{\cal S}({\bf x})~~~~~~~~~~{\rm for}~~B_r \ \ ,
&(4.7')\cr
E_r^-({\bf x}) &= a_{r-1}({\bf x}_\delta)
\,a_{r}({\bf x}_\delta)~~~~~~{\rm for}~~D_r \ \ , &(4.8')}
$$
where anyons with the cuts $\delta$ have been used.
They coincide with those given
in the ``bosonization formula''(3.15c).

This completes the proof that the deformed Lie algebras
${\cal U}_q(A_r)$, ${\cal U}_q(B_r)$
and ${\cal U}_q(D_r)$ are realized by the operators
$$
H_I=\sum_{{\bf x}\in \Omega} H_I({\bf x}) ~~~~~,
{}~~~~~E_I^\pm=\sum_{{\bf x}\in \Omega} E_I^\pm({\bf x})
\eqno(4.10)
$$
where the operators $H_I({\bf x})$ and
$E_I^\pm({\bf x})$ are defined with anyonic
oscillators according to eqs. (4.6-8) and (4.6'-8').

\vskip 3cm

%
%

\centerline{\bf 5. Anyonic Construction
of ${\cal U}_q(C_r)$}
\vskip 1cm

The anyonic realization of ${\cal U}_q(C_r)$ deserves
a special attention because the
Schwin- ger construction
of $C_r$ comes out naturally
in terms of bosonic oscillators and therefore involves
all the representations; instead the
discussion of section 3 shows that the realization of
a deformed Lie algebra by means of the ``bosonization
formula'' (3.15) makes use of the undeformed Lie algebra  in a
representation belonging to the set ${\Re}_{(0,1/2)}$.

To represent the algebra $C_r$ in terms of
{}fermionic oscillators, we have to embed it
into the algebra $A_{2r-1}$ [29]. By using
$2r$ fermionic oscillators
$c_{\alpha}({\bf x})$ for each point ${\bf x}$ of
the lattice, we write
$$
\eqalign{
e_i^+({\bf x}) &=
c_i^\dagger({\bf x})\,c_{i+1}({\bf x})
+c_{2r-i}^\dagger({\bf x})\,c_{2r-i+1}({\bf x})\ \ , \cr
e_i^-({\bf x}) &=
c_{i+1}^\dagger({\bf x})\,c_{i}({\bf x})
+c_{2r-i+1}^\dagger({\bf x})\,c_{2r-i}({\bf x})\ \ , \cr
h_i({\bf x}) &=
n_i({\bf x})-n_{i+1}({\bf x})
+n_{2r-i}({\bf x})-n_{2r-i+1}({\bf x})\ \ , \cr
}
\eqno(5.1)
$$
{}for $i=1,2,...,r-1$, in correspondence with the
short roots ${\alpha}_i$ of $C_r$, and
$$
\eqalign{
e_r^+({\bf x}) &=c_{r}^\dagger({\bf x})
\,c_{r+1}({\bf x})\ \ , \cr
e_r^-({\bf x}) &=c_{r+1}^\dagger({\bf x})
\,c_{r}({\bf x})\ \ , \cr
h_r({\bf x}) &=n_{r}({\bf x})-
n_{r+1}({\bf x})\ \ , \cr
}\eqno(5.2)
$$
{}for the long root $\alpha_r$ of $C_r$. It is easy
to check that the operators $h_I({\bf x})$,
$e_I^\pm({\bf x})$ defined in these equations
satisfy the commutation relations
(3.10) with the Cartan matrix appropriate for
$C_r$ (see Tab.1) and $q=1$. However for $i\not=r$ the square
of the operators $e_i^\pm({\bf x})$
does not vanish and therefore we
cannot immediately apply the
``bosonization formula'' (3.15) to construct the
$q$-deformation of $C_r$. In our fermionic
realization the fundamental representation
of $C_r$, which is characterized by the
Dynkin labels (1,0,...,0) of its highest weight,
acts on the $2r$-dimensional
vector space spanned by the states
$c_\alpha^\dagger({\bf x})|0\rangle$ with
$\alpha=1,2,...,2r$ (see Tab.3).
This representation obviously
belongs to the set ${\Re}_{(0,1/2)}$
since
the only eigenvalues of $h_I({\bf x})$ are 0 or
$\pm 1$ and the square of the operators
$e_I^\pm({\bf x})$ vanishes for $I=1,2,...,r$.

In order to select this representation we have to
impose a further condition on the
{}fermionic operators $c_{\alpha}({\bf x})$; we perform a
sort of Gutzwiller projection,
using hard-core fermions satisfying the extra condition
$$
c_\alpha({\bf x})\,c_\beta({\bf x})=
c_\alpha^\dagger({\bf x})\,c_\beta^\dagger({\bf x}) ~ = ~ 0
\eqno(5.3)
$$
{}for any $\alpha$, $\beta=1,2,..., 2r$.

We must also observe that we cannot deform
$C_r$ by simply replacing in eq.
(5.1) the fermionic oscillators with anyonic
ones defined as in (2.6-7). In fact, for $i\not =r$
the step operators constructed in this way
would not have the form (3.15b-c) as the disorder
operators contained in $a_i^\dagger\, a_{i+1}$ would
give an exponential different from those contained
in $a_{2r-i}^\dagger \,a_{2r-i+1}$.

This difficulty can be simply overcome by
requiring that the pair of anyons $a_I$ and
$a_{2r-I+1}$ $(I=1,2,...,r)$ arise from the
corresponding fermions
coupled to the same Chern-Simons field with
opposite charge.
Therefore the disorder operators to be used in
eq. (2.6) are
$$
K_I({\bf x}_{\alpha}) =
K_{2r-I+1}^\dagger({\bf x}_{\alpha}) =
{\rm \exp}{\Big[{\rm i} \,
\nu\sum\limits_{{{\bf y}\in \Omega}\atop {{\bf y}\not = {\bf x}}}
\Theta_{{{\alpha}_x}}({\bf x},{\bf y})\,
    \big(n_I({\bf y})- n_{2r-I+1}({\bf y})\big)}\Big]
\eqno(5.4)
$$
for $I=1,2,...,r$. The anyonic oscillators
defined in this way have the same generalized
commutation relations discussed in section 2,
and also non trivial
braiding relations between $a_I$ and $a_{2r-I+1}$,
{}for instance:
$$
a_I({\bf x}_\gamma)\,a_{2r-I+1}({\bf y}_\gamma)
+ q~a_{2r-I+1}({\bf y}_\gamma)\,a_I({\bf x}_\gamma) = 0
\, ~~~~~{\rm for}~~~{\bf x}>{\bf y}\ \  .
$$

With these definitions it is immediate to
check that eqs. (3.15) are reproduced
if
$$
\eqalign{
E_j^+({\bf x}) &=
a_j^\dagger({\bf x}_\gamma)\,a_{j+1}({\bf x}_\gamma)+
a_{2r-j}^\dagger({\bf x}_\gamma)\,
a_{2r-j+1}({\bf x}_\gamma)\ \ , \cr
E_j^-({\bf x}) &=
a_{j+1}^\dagger({\bf x}_\delta)\,a_{j}({\bf x}_\delta)
+a_{2r-j+1}^\dagger({\bf x}_\delta)\,
a_{2r-j}({\bf x}_\delta)\ \ , \cr
H_j~({\bf x}) &=
n_j({\bf x})-n_{j+1}({\bf x})+
n_{2r-j}({\bf x})-n_{2r-j+1}({\bf x})\ \ , \cr
}\eqno(5.5{\rm a})
$$
{}for $j=1,2,...,r-1$; and
$$
\eqalign{
E_r^+({\bf x}) &=
a_r^\dagger({\bf x}_\gamma)\,a_{r+1}
({\bf x}_\gamma)\ \ , \cr
E_r^-({\bf x}) &=
a_{r+1}^\dagger({\bf x}_\delta)\,a_{r}({\bf x}_\delta)\ \ , \cr
H_r({\bf x}) &=
n_r({\bf x})-n_{r+1}({\bf x})\ \ . \cr
}\eqno(5.5{\rm b})
$$

In these formulas, $n_i({\bf x})$ is given for
any value of $i$ by eq. (4.5) and
$q_r=q={\rm e}^{2{\rm i}\pi\nu}$
{}for the long root and $q_j=q^{1/2}$ for the
short roots ($j=1,2,...,r-1$).
The discussion of section 3 guarantees
therefore that the operators
$$
H_I=\sum_{{\bf x}\in \Omega} H_I({\bf x})~~~~,
{}~~~~E_I^\pm=\sum_{{\bf x}\in \Omega} E_I^\pm({\bf x})
$$
satisfy the generalized commutation relations
of the deformed algebra ${\cal U}_q(C_r)$.
\vskip 3cm
%
%
\centerline{\bf 6. Final Remarks}
\vskip 1cm
In this paper we have discussed the anyonic realization of
${\cal U}_q(G)$, being $G$ any classical Lie algebra.
{}For our construction it has been crucial the fact that the
{}fundamental representations of ${\cal U}_q(G)$ do not depend
on the deformation parameter $q$.
Therefore we believe that also ${\cal U}_q(E_6)$ and
${\cal U}_q(E_7)$
could be realized in terms of anyons, possibly introducing
a larger
number of them, analogously to the $q$-oscillator construction
of ref.[17].

The situation is instead quite different for ${\cal U}_q(E_8)$,
${\cal U}_q(F_4)$ and ${\cal U}_q(G_2)$, because their
{}fundamental representations are not in the class
${\Re}_{(0,1/2)}$.
Therefore these deformed algebras do not share the
{}fundamental representations with the undeformed ones.
This is in contrast with the possibility of building
anyonic realization of
${\cal U}_q(E_8)$, ${\cal U}_q(F_4)$ and ${\cal U}_q(G_2)$
of the type discussed in this paper. In fact their restriction
to a single site would be a fermionic representation
 no longer dependent on
the statistical parameter and therefore would be a representation of the
undeformed algebra.

The whole treatment of this paper can be extended to one
dimensional chains replacing the angles
$\Theta_{{\gamma}_x}({\bf x},{\bf y})$ and
$\tilde\Theta_{{\delta}_x}({\bf x},{\bf y})$ with $\pm{\pi\over 2}$
as specificated in eqs. (3.16).
In such a case it is also possible to assign real values
to the deformation parameter $q$, as in one dimension it
is no longer forced to be a pure phase.
Our construction is valid also in that case; in fact for
real $q$ all the equations of the paper still hold, once
that the ordering $\gamma$ and $\delta$ are exchanged
in the creation operators $a_i^\dagger({\bf x})$,
leaving unchanged the destruction operators
$a_i({\bf x})$.
The case of real $q$ is interesting because it leads
to unitary representations.

\vskip 3cm
%
%

\centerline{\bf Acknowledgments}
\vskip 1cm
\noindent
The authors would like to thank A. Sciarrino for an inspiring comment
and L. Castellani and A. Lerda for many useful discussions.
\vfill
\eject
%
%
\centerline{\bf References}
\vskip 1cm
\item{[1]}V.G. Drinfeld, {\it Sov. Math. Dokl.} {\bf 32}
(1985) 254;
\medskip
\item{[2]}M. Jimbo, {\it Lett. Math. Phys.} {\bf 10}
(1985) 63; {\bf 11} (1986) 247;
\medskip
\item{[3]} L.D. Faddeev, N. Yu. Reshetikhin and
 L.A. Takhtadzhyan , {\it Algebra and Analysis} {\bf 1} (1987) 178;
\medskip
\item{[4]}For reviews see for example: S. Majid,
{\it Int. J. Mod. Phys.} {\bf A5} (1990) 1;
P. Aschieri and L. Castellani, {\it ``An Introduction to
Non-Commutative Differential Geometry on Quantum Groups''}, Preprint
CERN-Th 6565/92, DFTT-22/92 to appear in {\it Int. J. Mod. Phys};
\medskip
\item{[5]}C. Zachos,{\it ``Paradigms of Quantum Algebras''},
Preprint ANL-HEP-PR-90-61;
\medskip
\item{[6]}J.M. Leinaas and J. Myrheim, {\it Nuovo Cim.}
{\bf 37B} (1977) 1;
\medskip
\item{[7]}F. Wilczek, {\it Phys. Rev. Lett.}
{\bf 48} (1982) 114;
\medskip
\item{[8]}F. Wilczek, in {\it Fractional Statistics
and Anyon Superconductivity} edited by F. Wilczek
(World Scientific Publishing Co., Singapore 1990);
\medskip
\item{[9]}For a review see for example:
A. Lerda, {\it Anyons: Quantum Mechanics of Particles
with Fractional Statistics} (Springer-Verlag, Berlin, Germany 1992);
\medskip
\item{[10]}A. Lerda and S. Sciuto, {\it ``Anyons and Quantum Groups''},
 Preprint DFTT 73/92, ITP-SB-92-73, to appear in {\it Nucl. Phys. }{\bf B};
\medskip
\item{[11]}J. Schwinger, in {\it Quantum Theory of
Angular Momentum} edited by L.C. Biedenharn and
H. Van Dam (Academic Press, New York, NY, USA 1965);
\medskip
\item{[12]}A. Macfarlane, {\it J. Phys.} {\bf A22}
(1989) 4581;
\medskip
\item{[13]}L.C. Biedenharn, {\it J. Phys.} {\bf A22}
(1989) L873;
\medskip
\item{[14]}C-P. Sun and H-C. Fu,  {\it J. Phys.}
 {\bf A22} (1989) L983;
\medskip
\item{[15]}T. Hayashi, {\it Comm. Math. Phys.} {\bf 127}
(1990) 129;
\medskip
\item{[16]}O.W. Greenberg, {\it Phys. Rev. Lett.} {\bf 64} (1990) 705;
\medskip
\item{[17]}L. Frappat, P. Sorba and A. Sciarrino, {\it J. Phys.}
{\bf A24} (1991) L179;
\medskip
\item{[18]}R. Caracciolo and M. R-Monteiro, {\it ``Anyonic Realization
of $SU_q(N)$ Quantum Algebra''} Preprint DFTT 05/93, to appear in
{\it Phys. Lett.} {\bf B};
\medskip
\item{[19]}P. Jordan and E. P. Wigner, {\it Z. Phys.} {\bf 47} (1928)
631; J. H. Lowenstein and J. A. Swieca, {\it Ann. of Phys.} {\bf 68}
(1971)172;
\medskip
\item{[20]}V.F. M\"{u}ller, {\it Z. Phys.}
{\bf C47} (1990) 301;
\medskip
\item{[21]}T. Curtright and C. Zachos, {\it Phys. Lett.} {\bf 243B}
(1990)237;
\medskip
\item{[22]} E. Fradkin, {\it Phys. Rev. Lett.}
{\bf 63} (1989) 322;
\medskip
\item{[23]}M. L\"{u}scher, {\it Nucl. Phys.}
{\bf B326} (1989) 557;
\medskip
\item{[24]}D. Eliezer and G.W. Semenoff,
{\it Phys. Lett.} {\bf 266B} (1991) 375;
\medskip
\item{[25]}D. Eliezer, G.W. Semenoff and S.S.C. Wu,
{\it Mod. Phys. Lett.} {\bf A7} (1992) 513;
\medskip
\item{[26]}D. Eliezer and G.W. Semenoff,
{\it Ann. Phys.} {\bf 217} (1992) 66;
\medskip
\item{[27]}L.P. Kadanov and H. Ceva, {\it Phys. Rev.} {\bf B3}
(1971) 3918;
E. Fradkin and L.P. Kadanov, {\it Nucl. Phys.} {\bf B170}
(1981) 1; E. Fradkin, {\it Field Theories of
Condensed Matter Systems} (Addison-Wesley, Reading, MA, USA 1991).
\medskip
\item{[28]}See for instance:
R. Slansky, {\it Phys. Rep.} {\bf 79} (1981) 1;
\medskip
\item{[29]} P. Goddard, W. Nahm, D. Olive and
A. Schwimmer, {\it Comm. Math. Phys.} {\bf 107} (1986) 179.

\vfill
\eject
\nopagenumbers
\centerline{{\bf Tab. 1 : Cartan Matrices of Simple Lie Algebras}}
\vskip1cm
$$
a\,[A_r] =
\left(\matrix{
 2&-1&0&0&.&.&0&0&0&0 \cr
 -1&2&-1&0&.&.&0&0&0&0\cr
 0&-1&2&-1&.&.&0&0&0&0\cr
 .&.&.&.&.&.&.&.&.&.\cr
 .&.&.&.&.&.&.&.&.&.\cr
 0&0&0&0&.&.&-1&2&-1&0\cr
 0&0&0&0&.&.&0&-1&2&-1\cr
 0&0&0&0&.&.&0&0&-1&2 }\right)
$$
\vskip1cm
$$
a\,[B_r] =
\left(\matrix{
 2&-1&0&0&.&.&0&0&0&0 \cr
 -1&2&-1&0&.&.&0&0&0&0\cr
 0&-1&2&-1&.&.&0&0&0&0\cr
 .&.&.&.&.&.&.&.&.&.\cr
 .&.&.&.&.&.&.&.&.&.\cr
 0&0&0&0&.&.&-1&2&-1&0\cr
 0&0&0&0&.&.&0&-1&2&-1\cr
 0&0&0&0&.&.&0&0&-2&2 }\right)
$$
\vskip1cm
$$
a\,[C_r] =
\left(\matrix{
 2&-1&0&0&.&.&0&0&0&0 \cr
 -1&2&-1&0&.&.&0&0&0&0\cr
 0&-1&2&-1&.&.&0&0&0&0\cr
 .&.&.&.&.&.&.&.&.&.\cr
 .&.&.&.&.&.&.&.&.&.\cr
 0&0&0&0&.&.&-1&2&-1&0\cr
 0&0&0&0&.&.&0&-1&2&-2\cr
 0&0&0&0&.&.&0&0&-1&2 }\right)
$$
\vskip1cm
$$
a\,[D_r] =
\left(\matrix{
 2&-1&0&0&.&.&0&0&0&0 \cr
 -1&2&-1&0&.&.&0&0&0&0\cr
 0&-1&2&-1&.&.&0&0&0&0\cr
 .&.&.&.&.&.&.&.&.&.\cr
 .&.&.&.&.&.&.&.&.&.\cr
 0&0&0&0&.&.&-1&2&-1&-1\cr
 0&0&0&0&.&.&0&-1&2&0\cr
 0&0&0&0&.&.&0&-1&0&2 }\right)
$$
\vfill
\eject
%
%
\nopagenumbers
\hskip 9cm \vbox{\hbox{DFTT 16/93}
\hbox{April 1993}}
\vskip 1.5cm
\centerline{\bf $q$-DEFORMED CLASSICAL LIE ALGEBRAS}
\centerline{{\bf AND THEIR ANYONIC
REALIZATION}
\footnote{${}^{\!*}$}{Work supported in part by
Ministero dell'Universit\`a e della Ricerca Scientifica
e Tecnologica.}}
\vskip 1cm
\centerline{{\bf Marialuisa Frau}${}^{\,a}$~,~
 {\bf Marco A. R-Monteiro}${}^{\,a,\,b,}$
\footnote{${}^{\!c}$}{Permanent address.}  ~~and~~
{\bf Stefano Sciuto}
\footnote{${}^{\!a}$}{\rm{e-mail addresses}:
sciuto(frau)@torino.infn.it,~~31890::sciuto(frau)}
${}^{,\,b}$}
\vskip 1.2cm
\centerline{\sl ${}^a$ Istituto Nazionale di Fisica Nucleare,
Sezione di Torino}
\vskip 0.50cm
\centerline{\sl ${}^b$ Dipartimento di Fisica Teorica, Universit\`a
di Torino}
\centerline{\sl Via P. Giuria 1, I-10125 Torino, Italy}
\vskip 0.50cm
\centerline{\sl ${}^c$ CBPF/CNPq, Rua Dr. Xavier Sigaud, 150 }
\centerline{\sl 22290 Rio de Janeiro, RJ, Brazil }
\vskip 2.5cm
\centerline{{\bf Abstract}}
\vskip 0.8cm

\noindent
All classical Lie algebras can be realized  \`a la Schwinger in terms
of fermionic oscillators. We show that the same can be done for their
$q$-deformed counterparts by simply replacing the fermionic oscillators
with anyonic ones defined on a two dimensional lattice. The deformation
parameter $q$ is a phase related to the anyonic statistical parameter.
A crucial r\^ole in this construction is played by a sort of bosonization
{}formula which gives the generators of the quantum algebras in terms of
the underformed ones.
The entire procedure works even on one dimensional chains; in such a case
$q$ can also be real.

\bye